# Spherical probes for simultaneous measurement of rotational and translational diffusion in 3 dimensions

Beybin Ilhan*, Jelle J. Schoppink, Frieder Mugele , Michael H.G. Duits

Physics of Complex Fluids, MESA+ Institute for Nanotechnology,

University of Twente, PO Box 217, 7500 AE Enschede, The Netherlands

**Abstract**

Real time visualization and tracking of colloidal particles with 3D resolution is essential for probing the local structure and dynamics in complex fluids. Although tracking translational motion of spherical colloids is well-known, accessing rotational dynamics of such particles remains a great challenge. Here, we report a novel approach of using fluorescently labeled raspberry-like colloids with an optical anisotropy to concurrently track translational and rotational dynamics in 3 dimensions. The raspberry-like particles are coated by a silica layer of adjustable thickness, which allows tuning the surface roughness. The synthesis and applicability of the proposed method is demonstrated by two types of probes: rough and smoothened.  The accuracy of measuring Mean Squared (Angular) Displacements are also demonstrated by  using these 2 probes dispersed in 2 different solvents. The presented 3D trackable colloids offer a high potential for wide range of applications and studies, such as probing crystallization dynamics, phase transitions and the effect of surface roughness on diffusion.

# Introduction

Studying colloidal dynamics via time-resolved locations of individual particles can provide microscopic insights to a variety of physical phenomena in different phases of matter[1-5]. Especially in systems with an intrinsic inhomogeneity, correlating the dynamics with the location inside the material can provide unique information that cannot be obtained with techniques that take an ensemble average over all particles, or any other bulk methods. Examples of research areas where local translational dynamics have been analyzed for this purpose include dense suspensions[6], glassy materials [7-9], polymer networks[10, 11], spatially confined materials[12-15], food products[16], cell biology [17, 18] and virus infection mechanisms[19].

In contrast to the many studies on translational dynamics, research related to rotational dynamics is rather scarce. This has been ascribed to a lack of experimental approaches for capturing rotational motion[20]. Rotational dynamics can shed a unique light onto various dynamic phenomena that cannot be accessed only with translational degrees of freedom, such as motion in glassy and supercooled states (where decoupling between translational and rotational diffusion emerges)[21-23]; particle adsorption and self-assembly at fluid interfaces [24, 25]; interfacial dynamics at solid-liquid interfaces[26] and biological interactions; such as viruses binding to membranes [27]

Ensemble averaged rotational diffusion of colloids has been studied by techniques such as fluorescence recovery after photobleaching (FRAP) [28, 29], depolarized dynamic light scattering[30, 31] and nuclear magnetic resonance (NMR) spectroscopy [32]. These bulk methods fall short in identifying local (dynamic) heterogeneities. Measuring rotational diffusion via individual probe particles can provide valuable local information in terms of dynamic length scales and structural signatures of complex fluids, such as local defects and crosslink densities in polymer networks[33], local rheology of soft materials[34], or intrinsic features of active fluids in biochemical processes[35].

In recent years, different strategies have been used towards tracking the rotational motion of diffusive probes of both spherical and anisotropic particle shapes such as rods, ellipsoids, and particle clusters[36-40]. Here, geometrical anisotropy is widely utilized because it naturally provides an identifiable optical axis to track angular displacements[20]. For spherical colloids, the lack of such a natural frame of reference requires a design where the optical isotropy is broken. A common type of such probes is the modulated optical nanoprobes (MOON) [41, 42]. This type of Janus particles usually consists of a fluorescent sphere that is half coated with a metal layer. Although these rotational probes are attracting interest in various fields [43], they have some drawbacks too. Due to the metal coating on one side, the surface chemistry is no longer uniform, and refractive index mismatches with the surrounding medium may compromises the image quality, especially for biological systems and high volume concentrations. Recently a new type of spherical rotational probes was introduced [23, 44]. These probes are bicolor or multicolor colloidal spheres with an eccentric core(s) shell structure, requiring multiple excitation wavelengths to be used. Although they provide homogenous surface chemistry, the centers of the core and the shell may not coincide precisely. In the case where multiple cores are utilized, they have to be overgrown to a rather large size, to attain a (near) spherical overall shape.

In this work, we introduce a novel and simple method for i) synthesizing fluorescently labelled raspberry-like spherical probes and ii) using them for simultaneously accessing rotational and translational dynamics in 3D. Our probes are made by densely covering a large silica core with many small $SiO_2$ particles, a fraction of which is fluorescently labeled. By coating these raspberries with a layer of silica of controllable thickness, we obtain particles of variable roughness while maintaining uniform surface chemistry. Optical anisotropy, that is introduced via the fluorescent tracers, allows for simultaneous tracking of the translational and rotational motion of each probe in 3D, using just one fluorescent dye. We demonstrate a proof of concept by tracking 2 types of probes (smoothened and rough) in the dilute regime.

## Results

**From Raspberries to Rotational Probes.** Raspberry particles are synthesized by coating the surface of positively functionalized SiO2 particles (*core*, r=1.04 ± 0.033 μm) with a dense layer of negatively charged small SiO2 particles (*berry*, r=0.164 ± 0.021 μm) via electrostatic heteroaggregation. To preserve mechanical integrity and to modify the surface roughness, a silica layer is overgrown on to these colloids via seeded growth. Figures 1b and 1c show confocal fluorescence images of typical probes with/out fluorescent berries. Figures 1d and 1e show SEM and TEM images for 2 systems differing in the layer thickness and hence in final size (they are identical otherwise): Surface rough probes (RP) with an rms roughness of 58 nm (4.5% relative to overall radius) and Smoothened probes (SP) with 23 nm rms (1.5% relative). (See Supporting Information for roughness characterization)

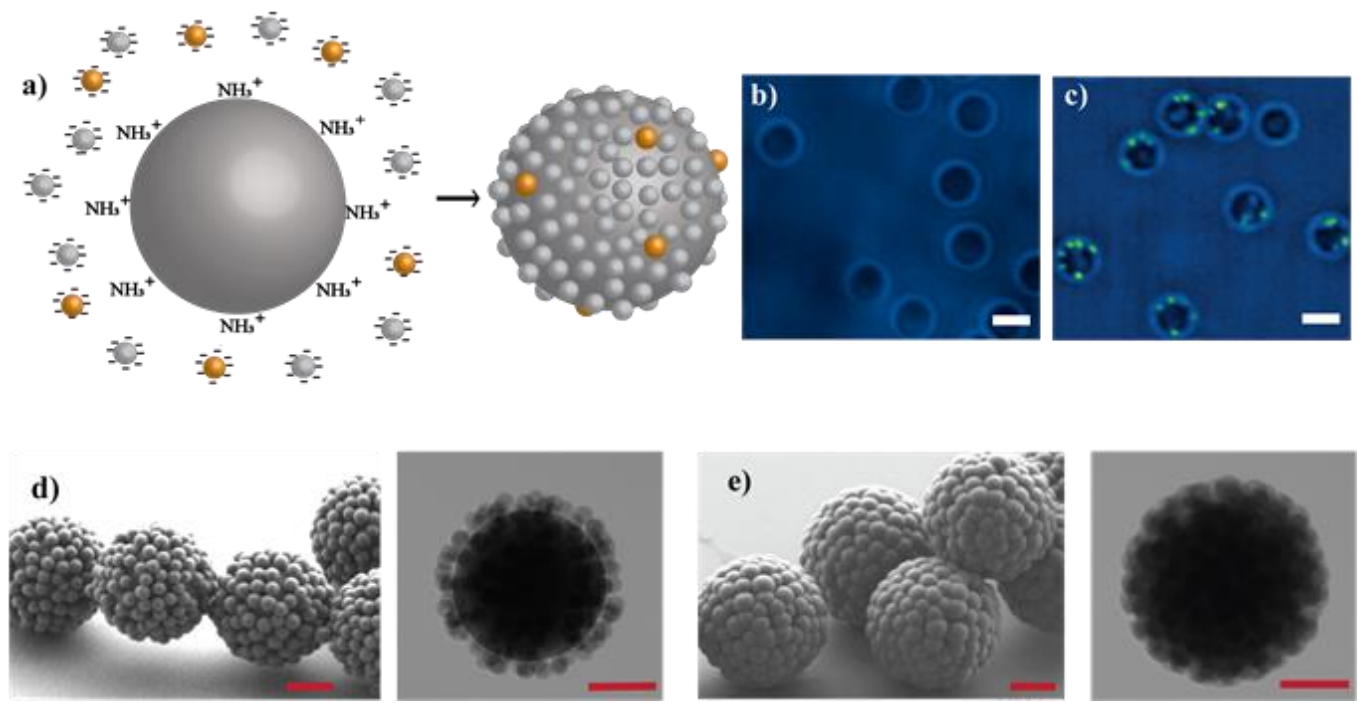

***Figure 1.*** *a) Illustration of the synthesis of the probe raspberries. b, c) superimposed brightfield and fluorescence image of raspberry probes without/with fluorescent berries(Scale bars are 2 μm). A movie of their Brownian motion can be seen in the Supporting Information as S.mov1. d, e) SEM and TEM images of rough (RP) and smoothened probes (SP). (Scale bars are 1 μm).*

**Extracting Rotation and Translation.** For optimal visualization, both type of probes are dispersed in refractive index matching solvent mixtures (n=1.45). Solvents are water-glycerol ($S_1$, 1:4 by weight, η=59 mPa.s) and water-glycerol-Dimethyl sulfoxide (DMSO) ($S_2$, 2:4:3 by wt., η=20 mPa.s). Particle volume fractions are chosen around 0.3%, to approach the dilute limit while keeping enough particles in the image volume for statistical analysis later on. Confocal Scanning Laser Microscopy (CSLM) in fluorescence mode is used to visualize only the labeled berries (see Supporting Information for details of experimental conditions). Their Cartesian coordinates are extracted using well-known particle locating algorithms[45, 46]. The located berries are then grouped in clusters to identify to which raspberry they belong; this is achieved using a maximum distance criterion (see Supporting Information for details). Only raspberries that contain 4 or more non-coplanar tracers are kept. This minimum number is required for simultaneously finding the center location (x,y,z) and optical radius ($R_{fit}$) of the raspberry particle, which is achieved via least-squares fitting to a sphere. Each obtained center location then provides an origin in a 3D Cartesian coordinate system that allows defining the spatial orientation of the raspberry probes based on the angular displacement of the tracer berries. In this scheme, the translation of each raspberry is extracted from the time-dependent center location, leaving the rotational motion to be measured from the changes in orientation. The latter is achieved using a modified algorithm, based on ref [38], by calculating angular displacements from a rotational transformation matrix in terms of 3 Euler angles. The key steps involved in dissecting translational and rotational motion are illustrated in Figure 2. Construction of trajectories from the time-dependent coordinates is achieved via publicly available tracking routines [46, 47]. The accuracy of the codes was tested with simulated data (mimicking typical experimental conditions) and gave good agreement (Figure S3, Supporting Information).

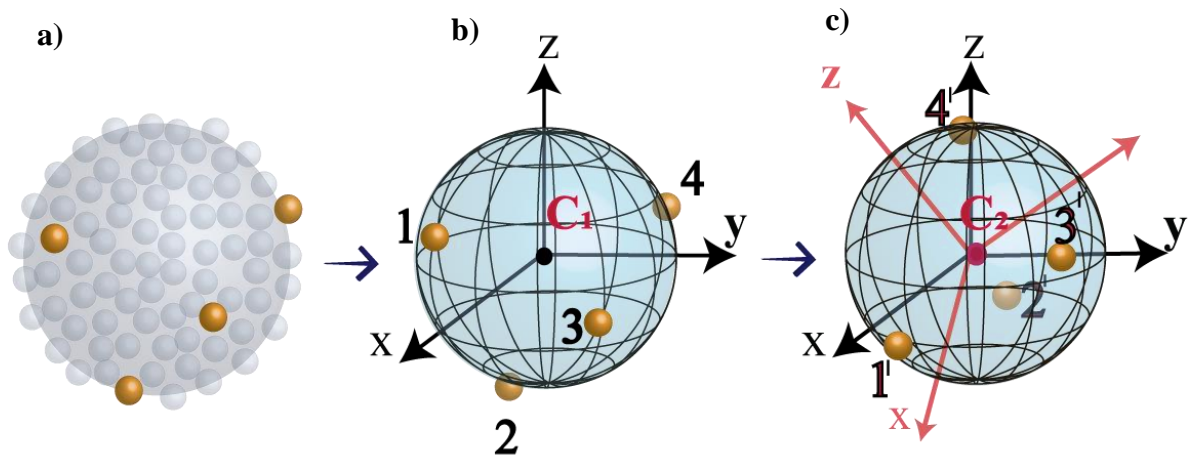

***Figure 2.*** *Key steps in measuring translation and rotation of a raspberry probe: a) 3D localization of the fluorescent berries, b) fitting a sphere that encompasses the cluster, c) tracking the rotational displacement using a frame of reference fitted at the core center.*

The coupled displacement of a group of berries on a raspberry is illustrated in Figure 3a for several time steps. Typical rotational and translational trajectories for a raspberry are shown in Figures 3b and c.

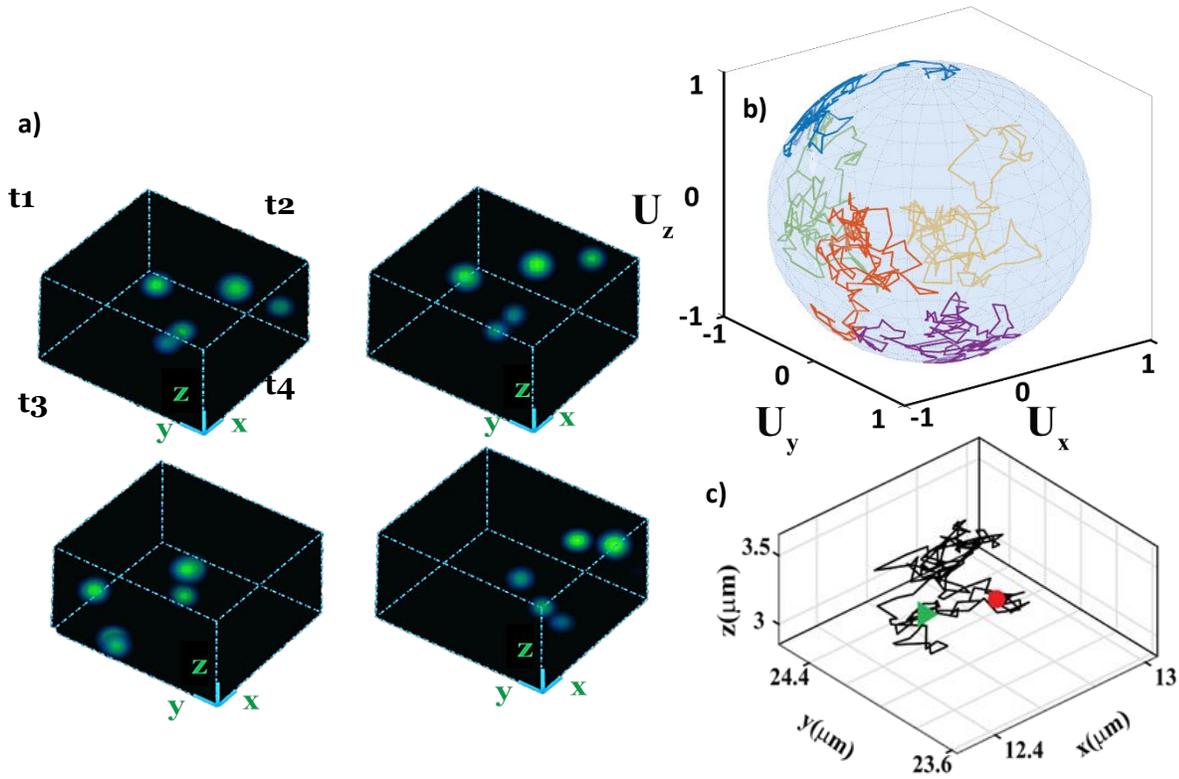

*Figure 3. Visual demonstration of typical results for tracking a single raspberry probe. a) Coupled motion of fluorescent berry particles forming a cluster at different time steps (3D rendering done with ImageJ FIJI), A movie can be seen in the Supporting Information as S.mov2, b) Orientational trajectories of the 5 berries (differing in color) projected onto the surface of a unit sphere, c) Center-of-mass trajectory of a single raspberry particle (green=begin, red=end),*

The mean squared displacement (MSD) versus lag time is calculated from the translational trajectories after the usual drift correction based on the ensemble averaged motion of the tracked particles. Hereafter the translational diffusion coefficient $D_t$ is calculated from:

$$\text{MSD} = 6D_t \Delta\tau \qquad (1)$$

The mean squared angular displacement (MSAD) is calculated in an analogous manner, but without the need for a drift correction. The rotational diffusion coefficient $D_r$ is obtained from equation (2) below (the detailed explanation of calculation from rotations around 3 principle axis is given in the experimental section and Supporting Information):

$$MSAD = 4D_r \Delta\tau \qquad (2)$$

**Validation of Brownian behavior.** We now examine the accuracy of measuring the diffusion coefficients for our raspberry probes, in different stages of the data analysis. First, we consider the fitted sphere's radius ($R_{fit}$) as extracted from the center locations of its berries. Given the dispersities of the core and berry systems, $R_{fit}$ should be close the sum of the typical radii: $R_{fit} \approx <R_{core}> + <R_{berry}>$, where the brackets indicate an average. Using transmission electron microscopy (TEM), we find $R_{core}$ = 1041 ± 33 nm and $R_{berry}$ = 160 ± 14 nm. Figure 4a shows $R_{fit}$ to be peaked at ≈1280 nm, giving a fairly close correspondence. The calculated standard deviation of 35 nm is the resultant of the two poly-dispersities and the typical uncertainty in $R_{fit}$, which is estimated to be 44 nm.

For the translational trajectories, the removal of drift is an essential correction, unless the lag time $\tau$ << $D_t/v^2$, with $v$ the drift velocity[48]. In our case, drift analysis also contributes to validation, because the vertical motion is dominated by sedimentation. The latter is illustrated in Figure 4b for a typical probe trajectory. Accordingly, the axial displacement of the ensemble of raspberries shows good linearity, as can be seen in Figure 4c. Since the particle volume fraction is very low, a comparison can be made between $v$ and theoretical Stokes sedimentation velocity of a smooth sphere. The dashed line in Figure 4c. is a linear fit that encompasses the axial drift displacement signal. Based on the Stokes Law, the effective radius of a corresponding smooth sphere is calculated from this linear fit is 1.28 μm ($\rho_p$=1.62±0.06 g/cm³), which is close to the average TEM radius. Drift signals in the horizontal directions are ascribed to the microscope translation stage.

The principal results of this work, obtained after all analysis steps and thus accumulating the inaccuracies of all these steps, are shown in Figures 4d and 4e. Here, the solid black lines represent the (drift corrected) MSD and the MSAD respectively. Both functions show a linear dependence on lag time, as expected for pure Brownian motion. The shortest addressed lag time is 2.4 s, as determined by the acquisition time of a 3D confocal scan. Both the MSD and the MSAD are significantly above their respective noise floors (which are shown in SI Figure. S4 b and c).

Because of the spherical symmetry of the system, none of the two diffusion coefficients should show any directionality. Decompositions of the MSD and MSAD along the X, Y and Z directions of the lab frame confirm this: the components along the different directions superimpose well in both Figure. 4d and Figure. 4e. Also the displacement histograms for a representative lag time (see insets) show a very good overlap and a Gaussian shape. This further corroborates that both types of diffusive motion are captured well.

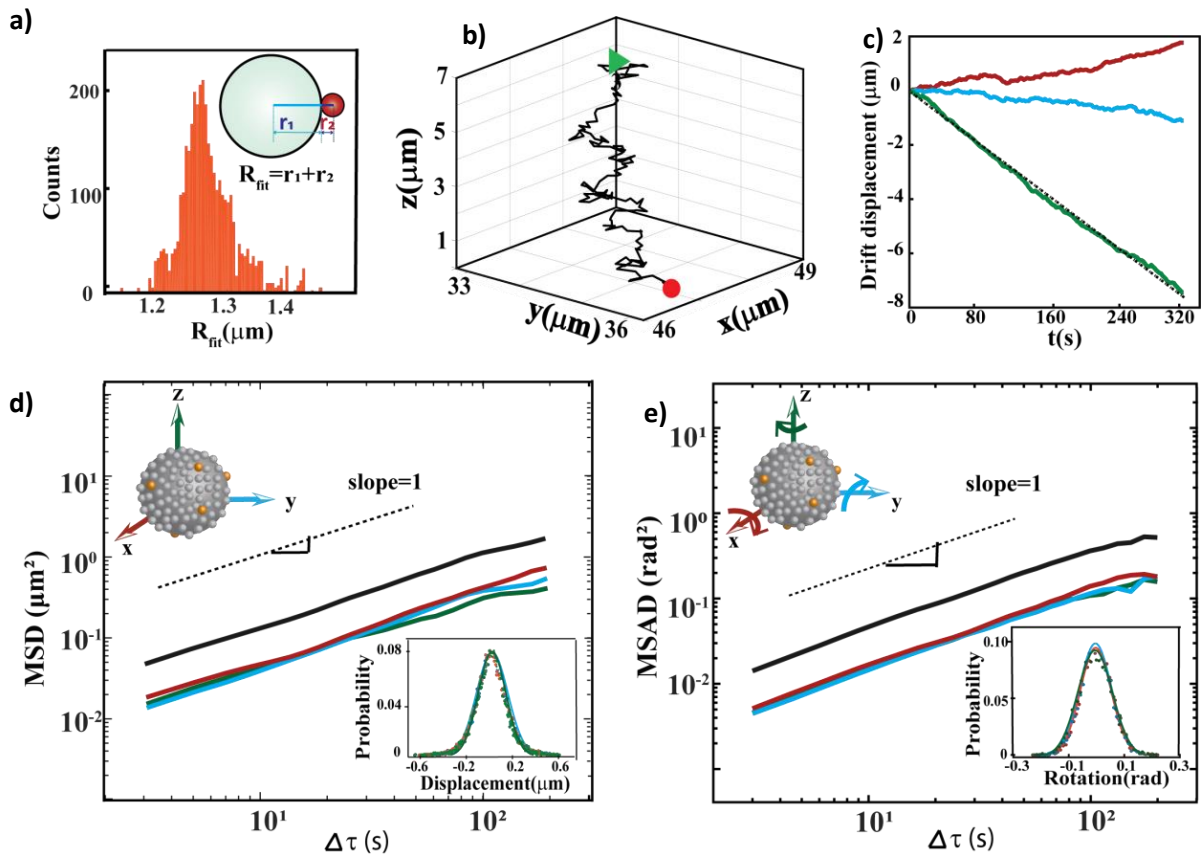

***Figure 4.*** *Key results of translational and rotational tracking. a) Histogram of fitted radius $R_{fit}$ (see inset) for both type of probes (which are optically identical). b) Typical 3D trajectory of a RP, before the removal of drift motion. c) Drift displacements for an ensemble of RP raspberries in solvent $S_1$, along the x, y and z directions. d) MSD and e) MSAD versus lag time (solid black lines) for RP probes in solvent $S_1$. Both functions are obtained from the same image data and contained 18 probes. Insets show decompositions along x, y and z for the MS(A)D, and for the displacement histograms. Color coding: red=x, blue=y, green=z.*

**Rotational and translational diffusion coefficients**. To assess the accuracy of the measured diffusion coefficients, we compare MSDs and MSADs for the two probe systems (RP, SP) in solvents with different viscosities: $S_1$ (59 ± 3 mPa.s) and $S_2$ (20 ± 1 mPa.s). Both probes are synthesized using the same core and berry particles, but for the SP system, a thicker silica layer has been overgrown to achieve a smoothened surface (Figure 1e). This thicker layer also gives a significant increase in the final probe size (Figure 5a). The MSAD measurements are shown in Figure 5b. Regardless of the solvent, both the RP and SP probes demonstrate a purely diffusive behavior, i.e. a linear increase in MSAD with lag time. Extracted numerical data are summarized in Table 1, where the diffusion coefficients are obtained from linear fits for lag times up to 10 s.

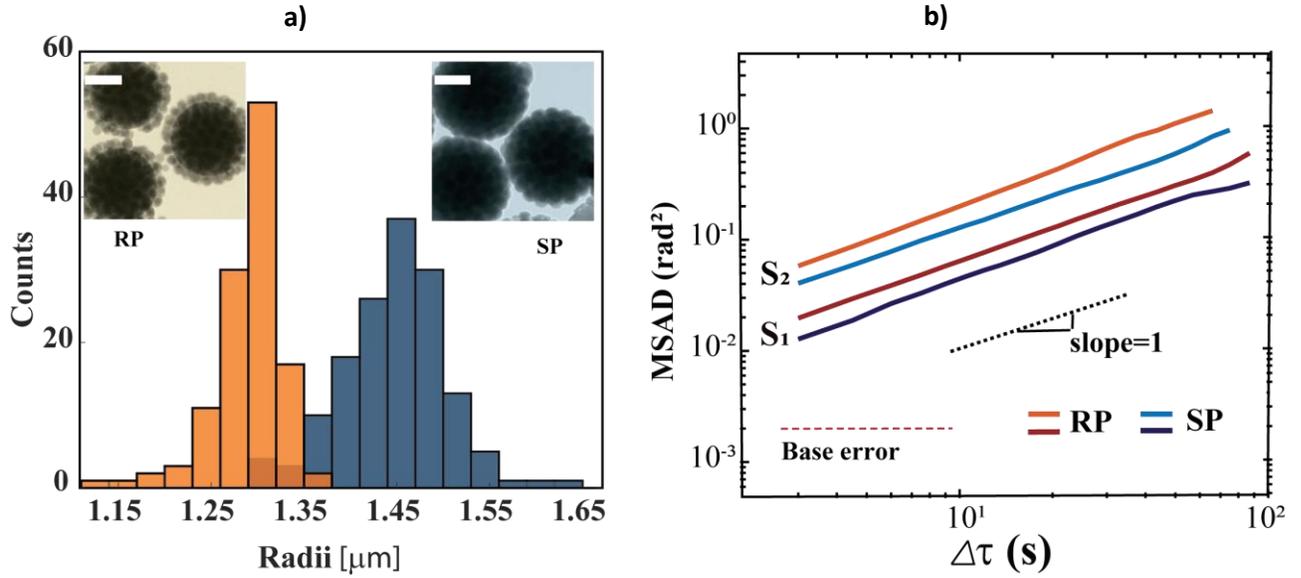

**Figure 5.** *a) Size distributions and TEM close-ups for RP (orange) and SP (blue). Scale bar is 1 μm. b) MSAD versus lag time for raspberry systems RP (orange/brown) and SP (blue/purple) dispersed in solvents $S_1$ (η=59 mPa.s) and in $S_2$ (η=20 mPa.s).*

**Table 1.** *Overview of measured diffusion coefficients and some derived quantities. $R_h$ is the hydrodynamic radius of the equivalent smooth sphere.*

|  | $D_t$ (μm²/s) | $D_r$ (s⁻¹) | $D_t / D_r$ (μm²) | $R_h [D_t]$ (μm) | $R_h [D_r]$ (μm) |
|---|---|---|---|---|---|
| **RP in $S_1$** | 2.78 10⁻³ | 1.58 10⁻³ | 1.76 | 1.29 | 1.20 |
| **RP in $S_2$** | 9.73 10⁻³ | 5.18 10⁻³ | 1.88 | 1.09 | 1.15 |
| **SP in $S_1$** | 2.23 10⁻³ | 1.13 10⁻³ | 1.97 | 1.61 | 1.34 |
| **SP in $S_2$** | 7.58 10⁻³ | 3.10 10⁻³ | 2.45 | 1.40 | 1.37 |

In the dilute limit, it is interesting to compare the measured $D_r$ and $D_t$ values to the theoretical expressions for smooth spheres in a Newtonian liquid with a no-slip boundary condition [49], $D_t$ is given by Stokes-Einstein equation and $D_r$ is given by Stokes-Einstein-Debye equation as follows:

$$D_t = \frac{k_b T}{6\pi\eta R} \quad (3)$$

$$D_r = \frac{k_b T}{8\pi\eta R^3} \quad (4)$$

where $k_b$ is the Boltzmann constant, T is temperature and R is radius. For rough spheres, $D_r$ and $D_t$ should be inversely proportional to the solvent viscosity η. Calculating $D_i(S_1)/D_i(S_2)$ with $i \in r, t$ for RP and SP separately, we obtain $η(S_1)/η(S_2)$=3.2 ± 0.3, in good agreement with the 3.0 ± 0.2 obtained from the measured viscosities.

Comparing the cases where the same probe system is dispersed in 2 different solvents, a different accuracy assessment can be made by calculating $D_t/D_r$. Now the viscosity effect is 'divided out' because of the inverse proportionality of both $D_t$ and $D_r$ to η. The results in Table 1 show very similar ratios (~1.82 ±0.08 μm$^2$) for RP and more differing ones (~2.2 ± 0.3 μm$^2$) for SP. The larger $D_t/D_r$ values for the SP probes are not unexpected, considering the $R^2$ proportionality for smooth spheres. The small variations among the measurements in different solvents might be due to a 'sampling effect': a combined effect of the poly-dispersity (Figure 5a) and the finite number of probes (typically 20-30) in a single image volume.

Lastly, we use the standard expressions (Eqns. 3 and 4) for $D_t$ and $D_r$ of smooth spheres in the dilute limit, to calculate effective radii $R_h$ from the measured diffusion coefficients and solvent viscosities. The $R_h$ values in Table 1 are comparable to those from the TEM measurements. Some slight differences between the values obtained from $S_1$ and $S_2$ are found; these might be because each of the 4 raspberry/solvent combinations was explored with a fresh solvent mixture (possibility of slight differences in viscosities) and a limited number of particles (possibly introducing a sampling effect). A striking observation is that for both RP and SP, the $R_h$ values obtained from $D_r$ correspond better with TEM measurements. Considering the $R^3$ proportionality of $D_r$ for smooth spheres (while it is ~$R^{-1}$ for $D_t$) it is likely that $D_r$ provides a more precise measure for $R_h$. Considering how close <$R_h$ [$D_t$]> and <$R_h$ [$D_r$]> are to those of the TEM measurements, we conclude that the surface roughness (being larger for RP) does not have a discernable effect on either of the two diffusion coefficients in the dilute limit.

**Application Scope .** Our novel probes can be useful in various cases where simultaneous 3D tracking of individual particle locations and orientations is needed. Due to tunability of the outer layer thickness, surface roughness can be altered. This offers a broad application potential for our particles in many colloidal dynamics studies. At low particle concentrations the effects of roughness on the two diffusion coefficients were too subtle to be measured. However at high concentrations, rough probes can be employed for studying the relation between roughness and diffusion. In the dense regime, particle-particle interactions play an important role in colloidal dynamics[50], and strong correlations between roughness and jamming have already been found[51-53]. In the smooth limit, our probes can be used for shedding more light on the effect of colloidal interactions on rotational diffusion [54, 55] or on transient phenomena like glassy dynamics and crystallization. Also in complex fluids whose structure is not dictated by particles, our (rough or smooth) probes could provide information about local (mechanical) properties. Here the surrounding 'bulk' material could e.g. be polymer solutions/gels or liquid interfaces.

Lastly, the synthesis and utilization of raspberry probes are not limited to the demonstrated methodology. Due to the simple synthesis and the use of only one fluorescent label for simultaneous tracking of two different motions, the 'berry platform' can pave the way for designing similar probes with additional shape isotropy. Applying the same concept while using different materials is yet another direction. Probes could be functionalized as active colloids, soft compressible particles, or serve as probes in bio-mimicking studies to resolve dynamics at biological processes[19, 56].

**Conclusions**. We developed a new type of colloidal probes with a near-spherical (i.e. raspberry) shape and homogenous surface chemistry. The use of fluorescent tracers in the shell allows to resolve both the center location and the orientation of each probe. As a proof of concept we demonstrated the simultaneous tracking of translational and rotational motion using time-resolved 3D confocal microscopy. The probes exhibit purely diffusive behavior in the dilute regime, with diffusion coefficients that are similar to theoretical values for smooth spheres of the same size. We envision that our probes can be employed in various cases where simultaneous 3D tracking of individual particle locations and orientations is needed.

**METHODS**

**Synthesis of Raspberry Probes with Optical Anisotropy**. Dense coverage of the cores by the berries is achieved via electrostatic hetero aggregation. While the presence of (+) amine groups on the cores, and (-) silanol ones on the berries already favors such aggregation, an optimization of the pH is needed to obtain interparticle bonds that are strong enough to prevent detachment by stirring forces. Meanwhile, also the stability (against homo-aggregation) has to be preserved for both cores and berries. Using $HNO_3$ to adjust the pH, Zeta potential measurements (with the Zetasizer) are conducted at varying pH for the aqueous dispersions of the particles. The results (shown in SI Figure S1) indicate an optimal pH of 4.5, where, $\zeta_{core}$=+31mV and $\zeta_{berry}$=−28mV, $\zeta_{fluo.berry}$=−21mV.

Besides the zeta potentials, also the mixing ratio of the cores and berries has to be considered. The number of berry particles needed to ensure dense coverage on a core particle is estimated by calculating how many berry particles can be fitted into the shell space between a hypothetical sphere with a radius of [$R_{core}$+2$R_{berry}$] and a core particle. For our system, this calculation gives 386 berries per core. In practice more berries are needed to ensure colloidal stability throughout the self-assembly process. Especially at the initial stages where the cores are only partially covered by berries, 'bridging aggregation' (berries binding to two cores) must be avoided. To prevent this a 10 times higher dose of berries is used. To obtain optical anisotropy, a mixture of fluorescent and plain berries is used. Assuming that all berries have an equal probability of being bonded to a core surface, the proportion of fluorescent berries is chosen to be ~ 1.0 %, to obtain on average 4 berries per core. The hetero-aggregation is achieved by adding cores to a suspension of berry particles (~1% wt) under mild stirring, and giving a reaction time of 1 hour. Excess berries are removed by 3 cycles of gravity settling and redispersion in aqueous solution at pH 4.5. A $SiO_2$ layer (thickness dependent on the type of the probe) is grown on these raspberry particles via seeded growth[57]. This layer serves two purposes: i) ensuring the mechanical integrity of the berries and ii) tuning the surface roughness from rough to smoothened.

**Rotational Tracking**. For the analysis of rotational displacements we followed an adapted methodology as described in ref. [38] As reported in detail before[38] orientation of the probes are determined from 3D rigid body transformations in between time steps. Since the center of mass locations of the probe particles are already obtained via sphere fitting, translational displacements are subtracted from the trajectories of the berry tracers so that the residual is only angular motion. Residual positional information of the berries is directly used to calculate rotational transformation matrix in between consecutive time steps.

$$\begin{bmatrix} x_i \\ y_i \\ z_i \end{bmatrix} = \mathbf{R} \begin{bmatrix} x_{i+1} \\ y_{i+1} \\ z_{i+1} \end{bmatrix} \qquad (5)$$

Where [x, y, z] coordinates denotes a location of an individual berry tracer at $i^{th}$ time step and **R** is the rotational transformation matrix. The calculation of R is as given in [38]. After obtaining this transformation matrix, we calculate the rotations around each principle axes. A general rotational matrix has a form of:

$$\mathbf{R} = \begin{bmatrix} \cos\beta\cos\gamma & \sin\alpha\sin\beta\cos\gamma - \cos\alpha\sin\gamma & \cos\alpha\sin\beta\cos\gamma + \sin\alpha\sin\gamma \\ \cos\beta\sin\gamma & \sin\alpha\sin\beta\sin\gamma + \cos\alpha\cos\gamma & \cos\alpha\sin\beta\sin\gamma - \sin\alpha\cos\gamma \\ -\sin\beta & \sin\alpha\cos\beta & \cos\alpha\cos\beta \end{bmatrix} \quad (6)$$

Where α, β and γ are the rotations around the principle X, Y, Z axes respectively. This rotation matrix is used to calculate the Euler angles directly[58] (see Supporting Information for details of extraction of angles). In a spherical system, due to the bounded nature of rotation angles, calculation of angular displacements relative to the axis of rotation will yield in displacements greater than a factor of 3/2 of the actual displacements[38, 59]. For that reason MSAD values are multiplied for a factor of 2/3 before calculating $D_r$ by using eq$^n$ (2).

## Supporting Information

**S.mov1.:** Brownian motion of Rough Probes dispersed in water and gravity settled onto a glass substrate.

**S.mov2.**: Reconstruction of the 3D motion of a typical RP dispersed in S1.

-Materials and Experimental conditions

-Zeta potential vs pH for the core and fluorescent berry particles.

-TEM size distributions for the core and berry particle systems.

-Detailed explanation of data analysis steps:

   **i)** Clustering algorithm

   **ii)** Extracting rotations in 3D

-Simulations

-Estimation of localization errors and noise floors for MSD and MSAD

**-Estimation of particle surface roughness:**

   **i)** AFM images of RP and SP particles

   **ii)** Example Illustration to sphere fit to AFM images for raspberry system RP and consecutive steps in roughness calculation